\documentclass[aps,pra,twocolumn,amsmath,amssymb,superscriptaddress,10pt]{revtex4-1}
\usepackage{amsfonts}
\usepackage{amsmath}
\usepackage{amssymb}
\usepackage{graphicx}
\usepackage[T1]{fontenc}
\usepackage[utf8]{inputenc}
\usepackage{hyperref}
\usepackage[dvipsnames]{xcolor}
\usepackage{color}
\usepackage{bm}
\usepackage{comment}
\usepackage{mathtools}
\usepackage[section]{placeins}
\usepackage[capitalize]{cleveref}
\usepackage{footnote}
\usepackage{etoolbox}
\usepackage{eqnarray}
\usepackage{pgfplots} 
\usetikzlibrary{plotmarks}
\pgfplotsset{compat=newest} 
\pgfplotsset{plot coordinates/math parser=false}

\begin{document}

\title{Analytic Form of a Two-Dimensional Critical Distribution}

\author{Steven T. Bramwell}
\affiliation{London Centre for Nanotechnology and Department of Physics and Astronomy, University College London, 17-19 Gordon Street, London WC1H 0AH, United Kingdom}

\begin{abstract}
This paper explores the possibility of establishing an analytic form of the distribution of the order parameter fluctuations in a two-dimensional critical spin wave model, or width fluctuations of a two dimensional Edwards-Wilkinson interface. It is shown that the characteristic function of the distribution can be expressed exactly as a Gamma function quotient, while a Charlier series, using the convolution of two Gumbel distributions as the kernel, converges to the exact result over a restricted domain. These results can also be extended to calculate the temperature dependence of the distribution and give an insight into the origin of Gumbel-like distributions in steady-state and equilibrium quantities that are not extreme values.
\end{abstract}

\maketitle

\twocolumngrid

Limit distributions play an important role in physical science, the best known being the Gaussian distribution, associated with the central limit theorem, and the three Fisher-Tippett distributions (Gumbel, Fr\'echet and Weibull) associated with extreme values~\cite{FT, book}. The Gumbel distribution, relevant to this paper, has probability density function (PDF) $g(x) = e^{-x-e^{-x}}$. Its application to extreme values is unambiguous, but more mysteriously, Gumbel and `Gumbel-like' distributions are also observed to apply to many quantities that are not extreme values.  These include quantities related to turbulence~\cite{BHP}, 
self-organised criticality~\cite{PRL}
$1/f$ noise~\cite{Racz}, 
river levels~\cite{EPL, Henrik},
electroconvection~\cite{Toth},
one-dimensional phonons~\cite{phonon},
glasses~\cite{Chamon},
plasmas~\cite{Milligan}, 
classical ~\cite{Joubaud} and quantum~\cite{Yakubo} phase transitions,  
porous media~\cite{Planet},  
radar signals~\cite{Barucci},  
galaxy distributions~\cite{Galaxy, Labini}, 
Kardar-Parisi-Zhang (`KPZ') surfaces~\cite{KPZ}
and atmospheric energy transfer~\cite{Blender} .   

Such observations define two distinct, but related, challenges. First, in what way (as seems likely) are these limit distributions, and second, what are the precise PDF's, how are they Gumbel-like and how do they arise from microscopic models?  To appreciate the importance of these challenges one only has to consider perhaps the oldest example of all:  the distribution of human lifetimes~\cite{Gompertz}. This obeys the Gompertz distribution -- mathematically a Gumbel~\cite{book} --  which has been described as one the `greatest quantitative laws of biology'~\cite{Shklovskii}. It would be a breakthrough indeed if a mechanistic explanation of this law could be found in terms of cellular structure or function~\cite{Kirkwood}. 

Condensed matter physics has something to offer here, as it provides examples of Gumbel distributions arising in one dimensional systems~\cite{phonon} or measures~\cite{Racz} and has inspired the discovery of some general principles relating Gumbel-like distributions to the asymptotic distribution of random sums~\cite{BertinClusel}.  In particular, as a higher-dimensional example of broad application~\cite{BHP,PRL,EPL,Toth,Milligan, Chamon, Joubaud, Yakubo, Planet, Barucci, KPZ}, much interest has centred on the width fluctuations of a two-dimensional Edwards-Wilkinson interface at steady state growth~\cite{EW, Racz_Plischke}, and equivalently, the equilibrium order parameter fluctuations of the two-dimensional XY model in the low temperature limit of its critical phase. These quantities share the same Gumbel-like distribution, which has sometimes been called the `BHP' distribution after the author and colleagues, who proposed its widespread relevance~\cite{BHP}. Its PDF, here called $\phi(z)$, has been approximated analytically as a generalised Gumbel $\sim g^{\pi/2}$~\cite{PRL} and has also been solved numerically~\cite{PRE, Palma}; but it has not been obtained in analytic form. The cumulants of $z$  may, however, be expressed exactly in terms of special functions~\cite{NPhys} and here this result is developed to calculate further properties of the PDF. This gives new insights into how
the appearance of a `real' Gumbel-like distribution can be understood in microscopic terms.     

The variable $z$ here is the order parameter for the XY model in its low temperature limit, shifted by its mean and scaled by its standard deviation (see below). It is a global quantity, found by summing over spin variables~\cite{PRE}, and so, at first sight, might have been expected to obey the central limit theorem, with Gaussian fluctuations in the thermodynamic limit. The central limit theorem will apply if individual variables are individually negligible and statistically independent. The non-Gaussian limit distribution arises because the spins of the XY model, though individually negligible, are strongly interacting; and when the spin system is diagonalised into normal modes, the new variables, though statistically independent, are not individually negligible~\cite{PRE}.

The variable $z$ in $\phi(z)$ has, by definition, zero mean and unit standard deviation. The PDF $\phi(z)$ is equivalent (in the present context) to any PDF $f(x)$ derived from it by replacing $z$ with $x = z \sigma + \mu$, where $\sigma, \mu$ are the standard deviation and mean of $x$ respectively. In what follows $f(x)$ shall be used to denote such PDFs,  and $\sigma, \mu$ will be allowed to take values that simplify the algebra. The calculated $f(x)$s are are transformed to 
$\phi(z)$ (or approximations to it) by rescaling the variable using $\mu$'s and $\sigma$'s calculated either analytically or numerically. Rather than repeatedly spell out this transformation, in what follows, it will simply be referred to as `rescaling the variable'.  

The independence of the variables in the normal mode description of the XY model is very convenient for analysis of the statistics of $z$ (or $x$), as its characteristic function can be obtained as an exact summation over modes~\cite{PRE}. Defining $t$ as the Fourier variable conjugate to $x$, the characteristic function has been shown to be:
\begin{widetext}
\begin{equation}\label{first}
\Phi(t)  =  \exp \left\{
-\sum_{\bf q \ne 0} 
(i t G_{\bf q}/N) -  (i/2) \arctan(t G_{\bf q}/N) 
+ \frac{1}{4} \log\left( 1 + t^2 G^2_{\bf q}/N^2\right)
\right\}.
\end{equation}
\end{widetext}
Here the summation is over $N = L^2$ modes of wave vectors 
${\bf q} = (q_x, q_y) =  (n_x, n_y)2 \pi/L$ where the integers $n_{x,y}$ each run from $-L/2$ to $L/2-1$ (with ${\bf q} = (0,0)$ excluded) and $G_{\bf q}  = 1/(2 -2 \cos(q_x)- 2\cos(q_y))$.  Inversion of the numerically summed equation by fast Fourier transform, followed by rescaling the variable, gives the probability density function $\phi(z)$.  In Fig. 1 it is confirmed that by  $L = 32$ the PDF is converged to the thermodynamic limit form and is indistinguishable from that found by using the quadratic approximation $G_{\bf q} = 1/q^2$. These results are in very accurate agreement with the previous numerical inversions of the characteristic function, e.g. Ref. \cite{PRE}, Fig.2 and Ref. \cite{Palma}, Fig.1 (where finite size corrections are identified). 

The cumulants of $x$ are the coefficients of the powers of $t$ in the formal expansion of $\log(\Phi)$ about $t = 0$. To calculate the thermodynamic limit PDF, it is convenient to use $G_{\bf q} = 1/q^2$, which eliminates factors of $N$, and also to rescale ${\bf q} \rightarrow (n_x, n_y)$  which will not affect the final PDF (the same symbols $x, t$ are retained). 
Then the cumulants of order $r > 1$ are~\cite{PRE, EPL}: 
\begin{equation}\label{sum}
\kappa_r = \frac{1}{2} \Gamma(r) \sum_{n_x, n_y \ne 0,0} \frac{1}{(n_x^2+n_y^2)^r},
\end{equation}
which already looks like a two-dimensional analogue of a similar sum (over integer $n$) for the cumulants $\gamma_r$ of the Gumbel distribution: 
\begin{equation}\label{gum}
\gamma_r = \frac{1}{2} \Gamma(r) \sum_{n \ne 0} \frac{1}{|n|^r}.
\end{equation}
The author previously noted \cite{NPhys} that, with the help of a historic result by G. H. Hardy (1919) and L. Lorenz (1871)~\cite{Glasser, Zu,Mc}, both summations can be exactly expressed in terms of special functions, with for $r > 1$:
\begin{equation}\label{relation}
\kappa_r= 2 \Gamma(r) \zeta(r) \beta(r) = 2 \gamma_r \beta(r),
\end{equation} 
where $\zeta$ is the Riemann zeta function and $\beta$ the Dirichlet beta function.  With increasing $r$, the function  $\beta(r)$ rapidly approaches unity, $\beta(r)\rightarrow 1$, resulting in $\kappa_r\rightarrow \eta_r$, where $\eta_r = 2 \Gamma(r) \zeta(r) = 2 \gamma_r$ represents the cumulants of the convolution of two Gumbel functions, $h(x) = g \ast g(x)$. 

\begin{figure}[!ht]
	\centering
	\includegraphics[width=1.0\linewidth]{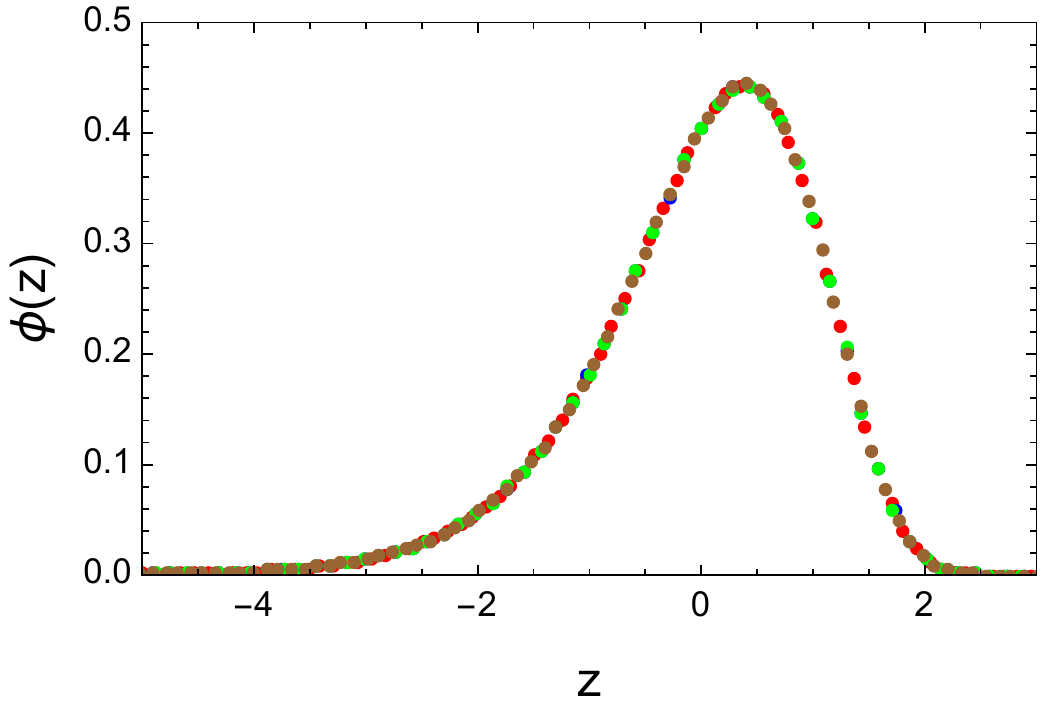}
	\includegraphics[width=1.0\linewidth]{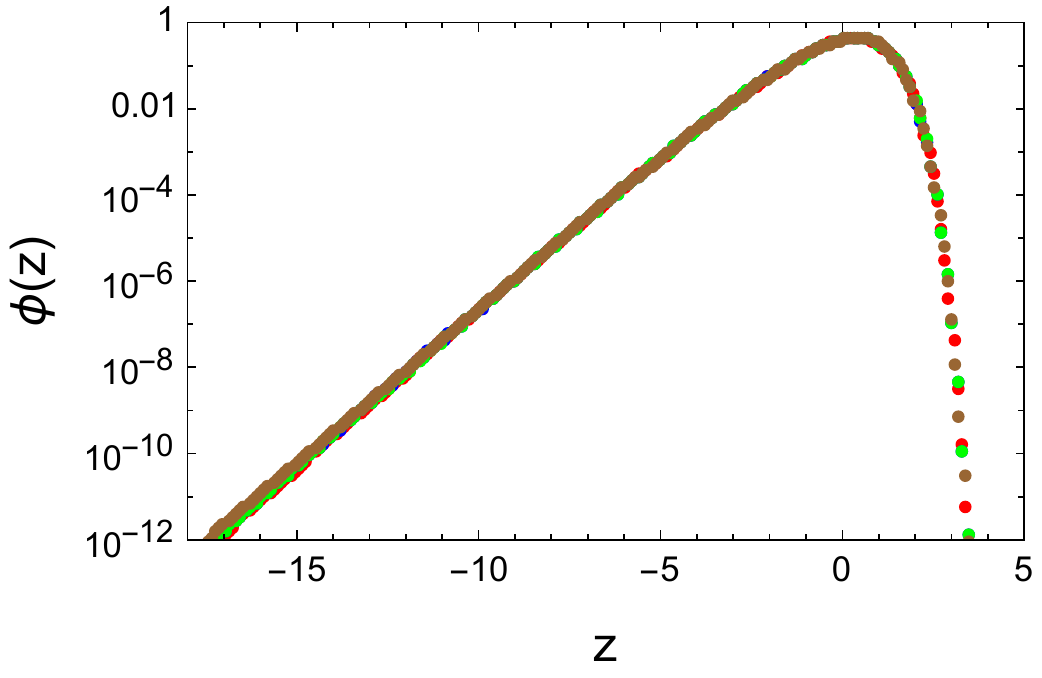}
	\caption{Close agreement between four different calculations of the limiting  PDF $\phi(z)$, illustrated on natural (upper) and logarithmic (lower) scales. The PDF is calculated by fast Fourier transform of the characteristic function $\Phi(t)$, expressed in four different ways. Blue circles (largely concealed by other symbols): using Eq. \ref{first} with `cosine' $G_{\bf q}$ and $L = 64$. Red circles: using Eq. \ref{first} with `quadratic' $G_{\bf q}$ and $L = 64$. Green circles: using Eq. \ref{first} with `cosine' $G_{\bf q}$ and $L = 32$. Brown circles: using Eq. \ref{exact} with 21 terms.  }
\end{figure}

The PDF $h(x)$, the convolution of two Gumbel functions, has been obtained analytically by Nadarajah \cite{Nadarajah}, with the result: 
\begin{equation}\label{Nad}
h(x)=2 e^{x } K_0\left(2 \sqrt{e^{x}}\right)
\end{equation}
where $K_0$ is a modified Bessel functions of the second kind. This PDF has mean $2\gamma$ where $\gamma$ is the Euler-Mascharoni constant, standard deviation $\sigma = \sqrt{2 \Gamma(2) \zeta(2)} = \pi/\sqrt{3}$ and  associated characteristic function $\Gamma(1 - i t)^2$. 

An exact analytic expression for the characteristic function $\Phi$ may now be obtained by representing the Dirichlet beta function in its series form, $\beta(r) = \sum_{n=0}^\infty(-1)^n/(2n+1)^r$. Using the cumulant series formed from Eq. \ref{relation}, this gives 
${\log \Phi=\sum_n(-1)^n\eta_r (i t/(2n+1))^r/r!)}$,  with the result
\begin{equation}\label{exact}
\Phi(t) = \prod_{n = 0, \infty} \Gamma(1- i t/( 2n +1))^{2(-1)^n},
\end{equation}
where the corresponding $f(x)$ has $\mu = \gamma \beta(1) = \gamma \pi/4$ and  
$\sigma = \sqrt{\kappa_2} = \sqrt{2 \Gamma(2) \zeta(2)\beta(2)}= \pi  \sqrt{\frac{{\rm Catalan}}{3}}$. Fast Fourier transform inversion of this expression, followed by rescaling of the variable gives near perfect agreement with the results of inverting Eqn. \ref{first}, as shown in Fig.1, but Eqn. \ref{exact} is far easier to compute. Indeed, as demonstrated in Fig, 2, the infinite quotient converges very quickly, with the first two approximations,
\begin{equation}\label{approx}
\Phi_0 = \Gamma[1- i t]^2, ~~~ \Phi_1 = \Gamma(1- it)^2/\Gamma(1- it/3)^2,
\end{equation}
already giving PDFs very close to the limiting form. Here $\Phi_0$ inverts to give $h(x)$ (Eqn. \ref{Nad}) and in Fig. 2 the analytic function is plotted. 

The facts that the PDF of interest is already well approximated by $h(x)$ and that
$\kappa_r$ rapidly converges on $\eta_r$ with increasing $r = 2,3,4\dots$, suggests that $f(x)$ may be profitably developed as a Charlier series, with $h(x)$ as the kernel. To simplify notation, define $\omega_r$ as the difference between the cumulants of $f(x)$ and $h(x)$:
\begin{equation}
\omega_r = \kappa_r - \eta_r = 2 \Gamma(r) \zeta(r) (\beta(r)-1), 
\end{equation}
and then the Charlier series is:
\begin{equation}
\begin{aligned}
f(x) = & \exp \left[ \sum_{r=1}^\infty \omega_r \frac{(-D)^r}{r!}\right] h(x) \\ 
= & \sum_{n=0}^\infty B_n\left(\omega_1, \omega_2 \dots,
\omega_n\right) \frac{(-D)^n}{n!} \, h(x).
\end{aligned}
\end{equation}
Here $D$ is the differential operator and the $B_n$s are complete Bell polynomials~\cite{Withers}. We have $B_0 = 1$ and can take $B_1 = 0$ because of the semi-invariant property of cumulants (i.e. the cumulants of order two or greater are invariant to a shift of the mean, so the mean of $f$ can be set equal to that of $h$). This gives: 
\begin{equation}\label{Charlier}
f(x) =  h(x) + 
\sum_{n=2}^\infty B_n\left( 0, \omega_2, \omega_3 \dots,
\omega_n \right) \frac{(-D)^n}{n!} \, h(x) 
\end{equation}
(see Appendix A for evaluation of this expression). 

\begin{figure}[!ht]
	\centering
	\includegraphics[width=1.0\linewidth]{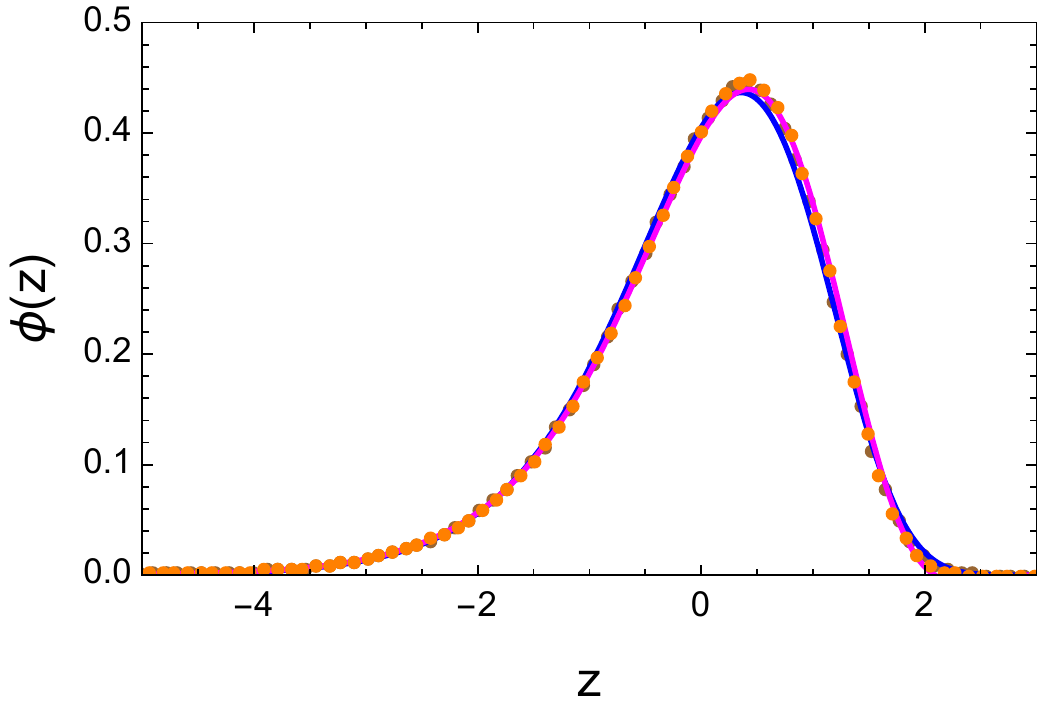}
	\includegraphics[width=1.0\linewidth]{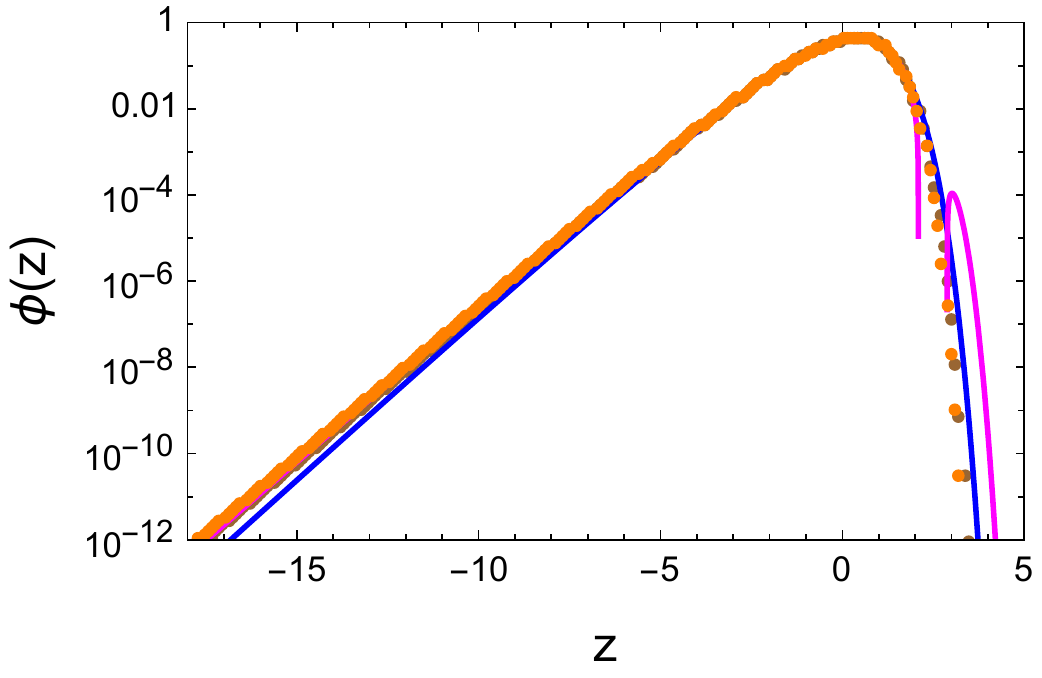}
	\caption{Comparison of approximations to the limiting PDF $\phi(z)$ on natural (upper) and logarithmic (lower) scales. Brown circles: the limiting form (same data as in Fig. 1). Orange circles: derived by fast Fourier transform from Eq. \ref{exact} with only 2 terms (i.e. $\Phi_1$ in Eq. \ref{approx}; these data largely conceal the brown circles). Blue line: $h(x)$, the convolution of two Gumbel functions (Eq. \ref{Nad}, conjugate to $\Phi_0$ in Eq. \ref{approx}). Magenta line: Charlier expression truncated to 6 terms (Eq. \ref{Charlier}; this goes negative around $z = 3$, where there is an apparent divergence of the series for $n > 24$ (not shown). 
	   }
\end{figure}

Eq. \ref{Charlier} is a simple expression to compute (e.g. in {\it Mathematica}~\cite{Math}) and to low order it appears to rapidly approach the limiting form (see Fig. 2). However for $n > 24$ the terms in the series rapidly grow larger and larger on the `steep' side of the PDF, moving well away from the limiting form.  Fig. 2 shows this apparent divergence already appearing when the series is truncated to six terms -- by 32 terms the effect is very severe (not shown). Nevertheless the result Eq. \ref{Charlier} does accurately converge to the (left hand) exponential tail of $\phi(z)$, as shown in Fig.2.

If the argument of $\exp(\log(\Phi(t)))$ is expanded in $it$ and truncated to any finite order, the Charlier method does return the exact Fourier transform of the truncated function. It can therefore be concluded that infinite summation of the series in $(i t)^n$ is essential to give an accurate approximation to $\phi(z)$. This is achieved exactly by the Eqn. \ref{exact} above, and also partially in the approximations $\Phi_0$ and $\Phi_1$ (Eq. \ref{approx}) where the series in $(i t)^n$ is partially summed to infinity by picking out contributions that define Gamma functions. Further insight into the failure of the truncated series is given in Appendix B. 

This calculated PDF $\phi(z)$ applies directly to the order parameter fluctuations of the XY model; to translate to `width-squared' fluctuations of an Edwards-Wilkinson interface at steady state growth~\cite{Racz_Plischke,PRE} $z$ needs to be replaced by $-z$ in $\phi(z)$ which reflects the PDF around $z=0$.

As mentioned in the introduction, there have been several analytic results for one dimensional problems that relate physical quantities to the Gumbel distribution $g(x)$: examples include the roughness of $1/f$ noise~\cite{Racz} and the one-dimensional phonon displacement distribution at zero temperature~\cite{phonon}. But analytic solutions in higher dimensional systems are much harder to come by. 
The cumulant expression  Eq. \ref{relation} and characteristic function, Eq. \ref{exact} are, to the author's knowledge, the only exact analytic results for such quantities relating to a critical distribution beyond one dimension.  A complete analytic form has still not been found, but these exact results suggest that one might yet be available in this two-dimensional system. 

Expressing the properties of the PDF in terms of analytic functions is potentially useful as it allows results to be generalised. To illustrate this principle, one could 
generalise to finite temperature in a Gaussian approximation of independent modes, by replacing $G_{\bf q} = 1/q^2$ with $G_{\bf q} = 1/q^{2-\eta}$ where $\eta = T/2\pi$ is the anomalous dimension (spin wave) exponent of the XY model. 
Since $\sum (q^{-2})^r = 4 \zeta(r)\beta(r)$ for $r >1$~\cite{Glasser}, it then follows that $\sum (q^{-(2-\eta)})^r = 4\zeta(r(1-\eta/2)) \beta(r(1-\eta/2))$. 
Thus the skewness $s$ of the distribution becomes:  
\begin{equation}\label{skew}
s(\eta) =\frac{2 \Gamma(3-3\eta/2 )\zeta(3-3\eta/2) \beta(3-3\eta/2)}{\left[2 \Gamma(2-\eta) \zeta(2-\eta) \beta(2-\eta)\right]^{3/2}}.
\end{equation}
With $\eta$ substituted for $T/2 \pi$, this expression is tested against existing numerical data~\cite{Banks} in Fig.3, where it is seen to accurately capture the temperature dependence of the skewness, as given by the (admittedly rather noisy) numerical data.   However, general arguments~\cite{Bruce} lead one to expect power law asymptotes for the critical PDF at finite temperature, more reminiscent of say the Weibull distribution than its limiting form, the Gumbel, and these are not likely to be captured by the Gaussian approximation. 

\begin{figure}[!ht]
	\centering
	\includegraphics[width=\linewidth]{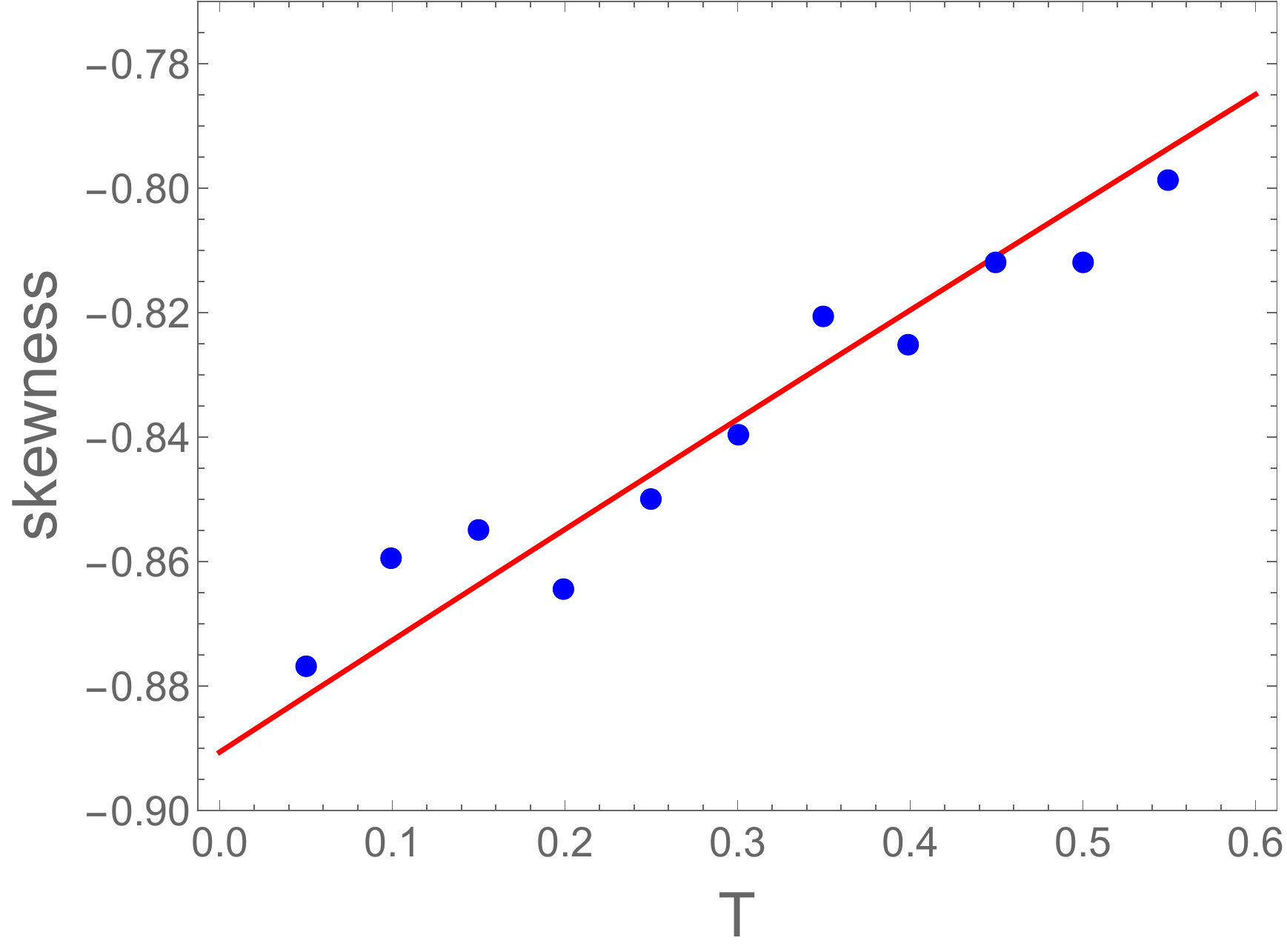}
	\caption{ Skewness of the PDF at finite temperature ($T$) calculated by Eq.\ref{skew} and compared to the numerical data of Ref. \cite{Banks} (system size $L=32$, which closely approximates the thermodynamic limit; see Fig. 1). }
\end{figure}

It is finally interesting to note that the function $\phi(z)$ is very close to the convolution of \underline{two} Gumbel distributions (see Eqn. \ref{relation} and Fig.2), with the underlined `two' seemingly related to the two-dimensionality of the problem.  That is, the two dimensional ($d=2$) mode structure with quadratic ($\delta = 2$) `dispersion' (Eq. \ref{sum}) can be closely approximated by two orthogonal one dimensional ($d=1$) mode structures with linear ($\delta = 1$) dispersions (Eq. \ref{gum}). Both of these mode structures are critical in the sense that the first moment diverges logarithmically with system size, approximating an integral 
($\int (1/q)\, dq$) in which all length scales ($\sim 1/q$) between the microscopic and macroscopic are equally important~\cite{PRE}. Thus, although the distribution $\phi(z)$, whose analytic form has been discussed here, is not exactly the convolution of two Gumbel functions, when viewed from the perspective of infinite lattice sums (Eqs. \ref{sum}, \ref{gum}), it is precisely the `two-dimensional equivalent' ($d = \delta = 2$) of the one-dimensional ($d=\delta = 1$) Gumbel distribution. Both of these distributions arise from a `critical' ($d = \delta$) structure of modes~\cite{EPL} --  a firm conclusion that is surely relevant to many of the real examples observed in physical systems.  

\begin{acknowledgments}
It is a pleasure to thank Michael Faulkner and Peter Holdsworth for useful comments and the anonymous referees for helping to improve the paper.   
\end{acknowledgments}

\appendix

\section{Evaluation of the Charlier series}

The first few complete Bell polynomials are: 
\begin{equation}
\begin{aligned}
& B_2(0,\omega_2) = \omega_2,\\
& B_3(0,\omega_2, \omega_3) = \omega_3,\\
& B_4(0,\omega_2,\omega_3, \omega_4) = 3 \omega_2^2 +\omega_4,\\
& B_5(0,\omega_2,\omega_3, \omega_4,\omega_5) =
10 \omega_3 \omega_2 + \omega_5,\\
& B_6(0,\omega_2,\omega_3, \omega_4,\omega_5, \omega_6) =
15 \omega_2^3 + 10 \omega_3^2 + 15 \omega_2 \omega_4 + \omega_6\\
\end{aligned}
\end{equation}
so that (Eq. \ref{Charlier}) 
\begin{equation}\label{fx}
\begin{aligned}
f(x) =  &h(x) + 
\frac{\omega_2 h^{(2)}(x) }{2!} -  \frac{\omega_3 h^{(3)}(x) }{3!}\\*
& +  \frac{(3 \omega_2^2 +\omega_4 )h^{(4)}(x) }{4!} 
+ \frac{(10 \omega_3 \omega_2 + \omega_5)h^{(5)}(x) }{5!}\\
& +\frac{(15 \omega_2^3 + 10 \omega_3^2 + 15 \omega_2 \omega_4 + \omega_6)h^{(6)}(x) }{6!}
+ \dots,
\end{aligned}
\end{equation}
 where $h^{(n)}$ is an nth derivative of $h(x)$. Or in terms of numbers: 
\begin{equation}\label{fxnum}
\begin{aligned}
f(x) =  &h(x)  -0.138231 h^{(2)}(x)  
+ 0.0248857 h^{(3)}(x)\\  
&+  0.00357113 h^{(4)}(x)
+ 0.00184635 h^{(5)}(x)\\
&+ 0.000250579 h^{(6)}(x)
+\dots.
\end{aligned}
\end{equation}

\section{Failure of of the truncated Charlier series}

In the derivation of Eq. \ref{exact}, $\beta(r)$ was expanded while $\zeta(r)$ was left intact. A different approach is to leave $\beta(r)$ intact and represent the zeta function by its series: $\zeta(r) = \sum_{k = 1}^\infty k^{-r}$. Summation of the cumulant series then gives:
\begin{equation}\label{B1}
\Phi(t) = \prod_{k = 1}^\infty{ \frac{\Gamma(3/4)^2 \Gamma(1/4- i t/4k)^2}{\Gamma(1/4)^2\Gamma(3/4- i t/4k)^2}}
\end{equation}
where a phase factor has been suppressed as it does not affect the final PDF.
Using a summation theorem \cite{WW}, this quotient can be expressed as the double product:
 \begin{equation}
\Phi(t) =\prod_{m= 1}^\infty  \prod_{k = 1}^\infty \frac{(m + 1/4)^2 (m+ 3/4-i t/4k)^2}{(m+ 3/4)^2(m + 1/4- i t/4k)^2}.
\end{equation}
For small $m, k$ this series can be Fourier transformed exactly to give, after rescaling the variable, approximations to $\phi(z)$, but this shows that, for any finite truncation, there is a singular point on the PDF where it hits zero on the `steep' side. This may be traced back to the fact that the original `gamma one half' variables that compose the distribution are strictly positive~\cite{PRE}. Taking the thermodynamic limit removes this singularity and allows the compound variables $x$ or $z$ to take unrestricted positive and negative values, with the resulting PDFs analytic everywhere on the real line.  The singularity is genuine in a finite system but it is removed by fast Fourier transform (Fig. 1) through a (rather arbitrary) discretisation. 

Without proving it in detail, it seems clear that the gamma functions (e.g. in Eq. \ref{exact} and \ref{B1}) represent thermodynamic limit summations, but any finite expansion of them restores the singularity. Hence the Charlier method, which expands all but the first term of Eq. \ref{exact}, fails when it meets the singularity.

\end{document}